\documentclass[conference]{IEEEtran}
\usepackage{epsfig}
\usepackage{times}
\usepackage{float}
\usepackage{afterpage}
\usepackage{amsmath}
\usepackage{amstext}
\usepackage{amssymb,bm}
\usepackage{latexsym}
\usepackage{color}
\usepackage{graphicx}
\usepackage{amsmath}
\usepackage{amsthm}
\usepackage{graphicx}
\usepackage[center]{caption}
\usepackage{pstricks}
\usepackage{caption}
\usepackage{subcaption}
\usepackage{booktabs}
\usepackage{multicol}
\usepackage{lipsum}
 \usepackage[T1]{fontenc}
\usepackage{epstopdf}

\allowdisplaybreaks

\newtheorem{thm}{Theorem}

\newtheorem{lem}{Lemma}

\newtheorem{rem}{Remark}
\newtheorem{example}{Example}
\theoremstyle{definition}

\providecommand{\definitionname}{Definition}

\newfloat{algorithm}{tbp}{loa}
\providecommand{\algorithmname}{Algorithm}
\floatname{algorithm}{\protect\algorithmname}

\long\def\comment#1{}

\newcommand{\dv}{{\mathbf d}}

\newcommand{\Zm}{{\mathbf Z}}

\newcommand{\Hc}{{\mathcal H}}

\newcommand{\Jc}{{\mathcal J}}

\newcommand{\Mc}{{\mathcal M}}

\newcommand{\Pc}{{\mathcal P}}
\newcommand{\Qc}{{\mathcal Q}}

\newcommand{\Sc}{{\mathcal S}}
\newcommand{\Tc}{{\mathcal T}}
\newcommand{\Uc}{{\mathcal U}}
\newcommand{\Wc}{{\mathcal W}}

\newcommand{\Xc}{{\mathcal X}}
\newcommand{\Yc}{{\mathcal Y}}

\newcommand{\rsf}{{\mathsf r}}

\newcommand{\Bsf}{{\mathsf B}}

\newcommand{\Gsf}{{\mathsf G}}
\newcommand{\Hsf}{{\mathsf H}}

\newcommand{\Ksf}{{\mathsf K}}

\newcommand{\Msf}{{\mathsf M}}
\newcommand{\Nsf}{{\mathsf N}}

\newcommand{\Rsf}{{\mathsf R}}

\newcommand{\Xsf}{{\mathsf X}}
\newcommand{\Ysf}{{\mathsf Y}}
\newcommand{\Zsf}{{\mathsf Z}}

\renewcommand{\arg}{{\hbox{arg}}}

\newcommand{\be}{\begin{equation}}
\newcommand{\ee}{\end{equation}}
\newcommand{\bea}{\begin{eqnarray}}
\newcommand{\eea}{\end{eqnarray}}

\begin{document}

\title{On the Benefits of Asymmetric Coded Cache Placement in Combination Networks with End-User Caches}

\author{
\IEEEauthorblockN{%
Kai~Wan\IEEEauthorrefmark{1}, %
Mingyue~Ji\IEEEauthorrefmark{2}, %
Pablo~Piantanida\IEEEauthorrefmark{1}, %
Daniela~Tuninetti\IEEEauthorrefmark{3}%
}
\IEEEauthorblockA{\IEEEauthorrefmark{1}L2S CentraleSupélec-CNRS-Université Paris-Sud, 
France, \{kai.wan, pablo.piantanida\}@l2s.centralesupelec.fr}%
\IEEEauthorblockA{\IEEEauthorrefmark{2}University of Utah, Salt Lake City, 
USA,  mingyue.ji@utah.edu}%
\IEEEauthorblockA{\IEEEauthorrefmark{3}University of Illinois at Chicago, Chicago, 
USA, danielat@uic.edu}%
}

\maketitle

\begin{abstract}
\emph{THIS PAPER IS ELIGIBLE FOR THE STUDENT PAPER AWARD.} This paper investigates the fundamental tradeoff between cache size and download time 
in the {\it $(\Hsf,\rsf,\Msf,\Nsf)$ combination network},
where a server with $\Nsf$ files 
is connected to $\Hsf$ relays (without caches) 
and each of the $\Ksf := \binom{\Hsf}{\rsf}$ users (with caches of size $\Msf$ files) is connected to a different subset of  $\rsf$ relays. 
Existing schemes fall within two categories: either use the uncoded symmetric cache placement originally proposed for the shared-link model and design delivery phase dependent on the network topology, or effectively divide the combination network into $\Hsf$ independent shared-link networks each serving $\Ksf^\prime := \binom{\Hsf-1}{\rsf-1}$ users; in either case, the placement phase is independent of network topology.
In this paper, a novel strategy is proposed where the {\it coded cache placement is dependent on network topology}. 
The proposed scheme is shown to be information theoretically optimal for large cache size.
In addition, when not exactly optimal, the proposed scheme can also outperform existing schemes. 
\end{abstract}

\section{Introduction}
\label{sec:intro}
Caching content at the end-user's memories smooth the network traffic. A caching scheme comprises two phases. 
(i) {\it Placement phase}: during off-peak hours, the server places parts of its library into the users' caches without knowledge of what the users will later demand. 
When pieces of files are simply copied into the cache, the cache placement phase is said to be {\it uncoded}; otherwise it is {\it coded}.
(ii) {\it Delivery phase}: each user requests one file during peak-hour time.  According to the user demands and cache contents, the server transmits the smallest number of files in order to satisfy the user demands.  

Caching was originally studied by Maddah-Ali and Niesen (MAN) in~\cite{dvbt2fundamental} for {\it shared-link} networks, which comprises a server with $\Nsf$ files, $\Ksf$ users with a cache of size $\Msf$ files, and an error-free broadcast link. The MAN scheme uses uncoded cache placement and a binary linear network code to deliver coded messages that are simultaneously useful for $t+1 := \Ksf \Msf/\Nsf+1$ users. Coded caching was shown to provide a multiplicative {\it coded caching/multicast gain} of $t+1$ over conventional uncoded caching schemes.
In~\cite{yas2}, a variation of the MAN scheme was shown to be information theoretically optimal to within a factor $2$ for shared-link networks.

\begin{figure}
\centerline{\includegraphics[scale=0.16]{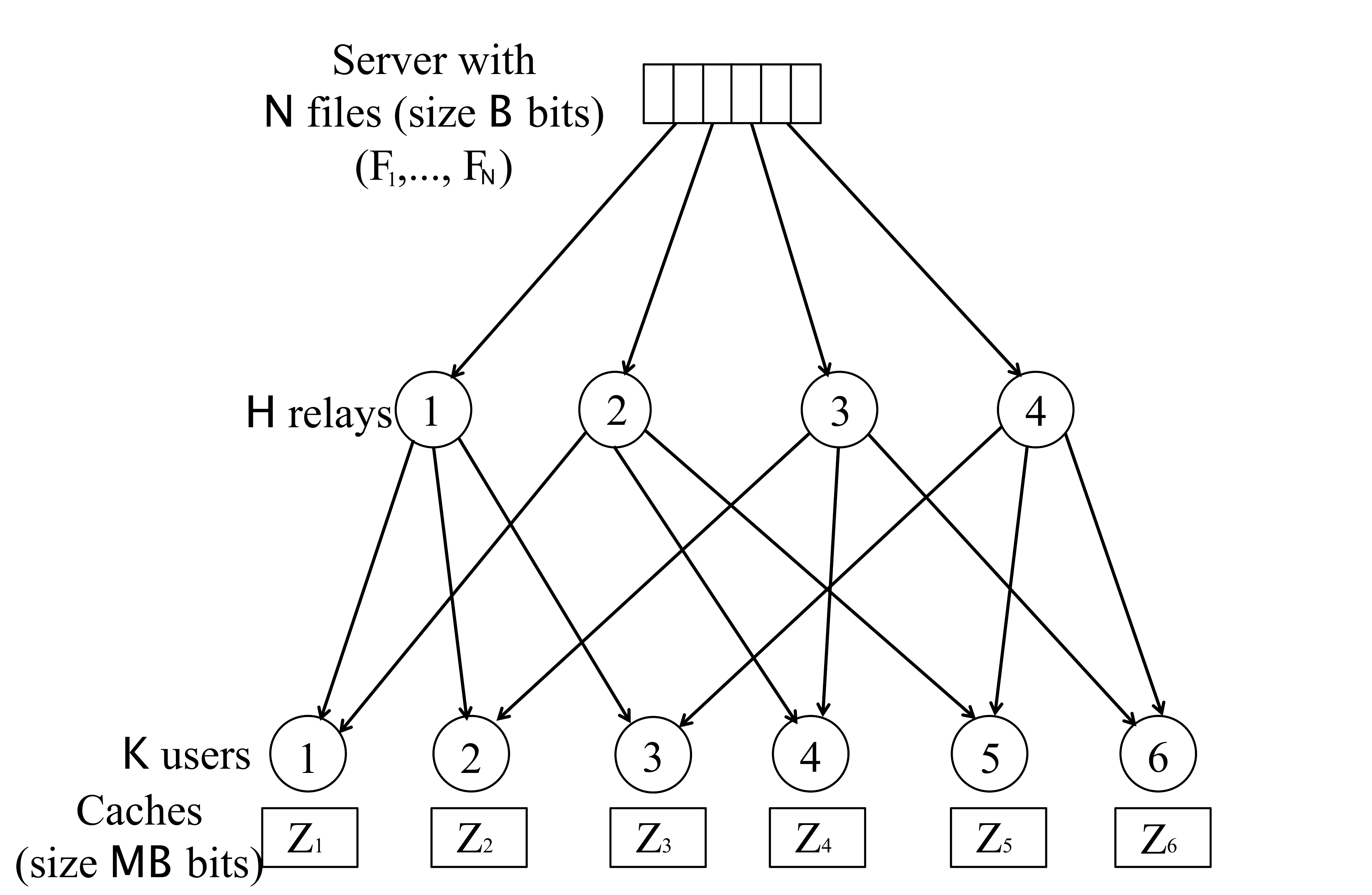}}
\caption{\small A combination network with end-user-caches, with $\Hsf=4$ relays and $\Ksf=6$ users, i.e., $\rsf=2$.}
\label{fig: Combination_Networks}
\vspace{-5mm}
\end{figure}

Since users may communicate with the central server through intermediate relays, caching in relay networks has recently been considered. 
Since it is difficult to analyze general relay networks, a symmetric network, known as {\it combination network}~\cite{cachingincom}, has received a significant attention. A $(\Hsf,\rsf,\Msf,\Nsf)$ combination network comprises a server with $\Nsf$ files that is connected to $\Hsf$ relays (without caches) through $\Hsf$ orthogonal links, and each of the $\Ksf := \binom{\Hsf}{\rsf}$ users (with caches of size $\Msf$ files) is connected to a different subset of  $\rsf$ relays through $\rsf$ orthogonal links, as shown in Fig.~\ref{fig: Combination_Networks}. 

\paragraph*{\textbf{Past Work (for combination networks)}}
Existing works use MAN uncoded placement for shared-link networks for the placement phase (which is agnostic of the network topology) and then design the delivery phase by leveraging the network topology~\cite{cachingincom,novelwan2017,wan2017novelmulticase}; these schemes are {\it symmetric} in the sense that for every file there exists one subfile cached by each subset of $t:= \Ksf \Msf/\Nsf$ users.
The main limitation of the MAN placement is that 
the multicasting opportunities (directly related to the overall coded cahing gain) to transmit the various subfiles are not ``symmetric'' across subfiles (because relays are connected to different sets of users).
One way to deal with this limitation is to divided the combination network into $\Hsf$ independent shared-link network and to precode every file by an MDS  (Maximum Distance Separable) code so that it becomes irrelevant from which relay a user has received a coded subfile--as long as enough coded subfiles have been collected~\cite{Zewail2017codedcaching}. The limitation of this coded placement is that the coded caching gain is now that of a network with $\Ksf^\prime := \binom{\Hsf-1}{\rsf-1} < \Ksf$ equivalent users, which appears to be suboptimal in light of known results for shared-link networks (i.e., the coded caching gain {\it fundamentally} scales linearly with the number of users $\Ksf$).

\paragraph*{\textbf{Contributions}}
In this paper we propose a novel placement that aims to attain identical ``multicasting opportunities'' for each coded subfile, which is then delivered by using a variation of the scheme proposed in~\cite{wan2017novelmulticase}.
Interestingly, our asymmetric placement leads to a ``symmetric delivery''--to be made precise later.
The novel scheme is proved to be {\it information theoretically optimal} 
when $\Msf\geq \frac{(\Ksf-\Hsf+\rsf-1)\Nsf}{\Ksf}$.  
To the best of our knowledge, this is the very first work that characterizes the exact memory-download time tradeoff for combination networks. In addition, when not optimal, the proposed scheme can also outperform  state-of-the-art schemes.

The paper is organized as follows.
Section~\ref{sec:model} gives the formal problem definition and states the main results.
Section~\ref{sec:proof of thm 1} contains proofs and numerical evaluations.

\section{System Model and Main Results} 
\label{sec:model}

We use the following notation convention. A collection is a set of sets, e.g., $\big\{\{1,2\},\{1,3\} \big\}$.
Calligraphic symbols denote sets or collections, 
bold symbols denote vectors, 
and sans-serif symbols denote system parameters.
We use $|\cdot|$ to represent the cardinality of a set or the length of a vector;
$[a:b]:=\left\{ a,a+1,\ldots,b\right\}$ and $[n] := [1:n]$; 
$\oplus$ represents bit-wise XOR. 
We define the set $\arg \max_{x\in \Xc}f(x) := \big\{x\in \Xc:f(x)=\max_{x\in\Xc}f(x)\big\}.$
Our convention is that $\binom{x}{y}=0$ if $x<0$ or $y<0$ or $x<y$. 

\subsection{System Model}
\label{sub:system model}
In a $(\Hsf,\rsf,\Msf,\Nsf)$ combination network, a server has $\Nsf$ files, denoted by $F_1, \cdots, F_\Nsf$, each composed of $\Bsf$ i.i.d uniformly distributed bits.
The server is connected to $\Hsf$ relays through $\Hsf$ error-free orthogonal links. 
The relays are connected to $\Ksf := \binom{\Hsf}{\rsf}$ users 
through $\rsf \, \Ksf$ error-free orthogonal links.
Each user has a local cache of size $\Msf\Bsf$ bits, for $\Msf\in[0,\Nsf]$,
and is connected to a distinct subset of $\rsf$ relays. 

The set of users connected to  relay $h$ is denoted by $\Uc_{h}, \ h\in[\Hsf]$. 
The set of relays connected to user $k$ is denoted by $\Hc_{k}, k\in[\Ksf]$.
For each set of relays $\Jc\subseteq[\Hsf]$,
we denote the common connected users for the relays in $\Jc$ by $\Pc_{\Jc}$.
For the network in~Fig.~\ref{fig: Combination_Networks}, for example, 
$\Uc_{1}=\{1,2,3\}$,
$\Hc_{1}=\{1,2\}$  and $\Pc_{\{1,2\}}=\{1\}$.

In the placement phase, user $k\in[\Ksf]$ stores information about the $\Nsf$ files in its cache of size $\mathsf{MB}$ bits, where $\Msf \in[0,\Nsf]$.  
The cache content of user $k\in[\Ksf]$ is denoted by $Z_{k}$; let $\Zm:=(Z_{1},\ldots,Z_{\Ksf})$.
During the delivery phase, user $k\in[\Ksf]$ requests file $d_{k}\in[\Nsf]$;
the demand vector $\dv:=(d_{1},\ldots,d_{\Ksf})$ is revealed to all nodes. 
Given $(\dv,\Zm)$, the server sends a message $X_{h}$ 
of $\Bsf \, \Rsf_{h}(\dv,\Zm)$ bits to relay $h\in [\Hsf]$. 
Then, relay $h\in [\Hsf]$ transmits a message $X_{h\to k}$ 
of $\Bsf \, \Rsf_{h\to k}(\dv,\Zm)$ bits to user $k \in \Uc_h$. 
User $k\in[\Ksf]$ must recover its desired file $F_{d_{k}}$ from $Z_{k}$ and $(X_{h\to k} : h\in \Hc_k)$ with high probability when $\Bsf\to \infty$. 
The objective is to determine the optimal {\it max-link load} defined as 
\begin{align}
\Rsf^{\star}
:=
\min_{\substack{\Zm}}\negmedspace
\max_{\substack{k\in\Uc_h, h\in[\Hsf],\\ \dv\in[\Nsf]^{\Ksf}}} \negmedspace
\max 
\left\{
\Rsf_h(\dv,\Zm),
\Rsf_{h\to k}(\dv,\Zm)
\right\}.
\end{align}
Since the max-link load of the uncoded routing scheme in~\cite{cachingincom} is $\Rsf_{\rsf}=\Ksf/\Hsf (1-\Msf/\Nsf)$, we define the {\it coded caching gain} $g$ of a scheme with max-link load $\Rsf$ as 
\begin{align}
g:=\frac{\Rsf_{\rsf}}{\Rsf}=\frac{\Ksf/\Hsf (1-\Msf/\Nsf)}{\Rsf}. 
\end{align}
Define that $\Ksf^{\prime}:=\binom{\Hsf-1}{\rsf-1}$, where $\Ksf^\prime$ is the number of users connected to each relay. By the cut-set bound in~\cite{cachingincom}, $g\leq \Ksf^{\prime}$.

\subsection{Main Results}
\label{sub:mainthms}
We now state our main results. 
Thm.\ref{thm:load of r>2 g general} gives the  max-link load of the novel proposed scheme with coded asymmetric cache placement and Thm.\ref{thm:optimaity large memory size} gives the optimality results. 
 Different from the state-of-the-art schemes, which fix the cache size and compute the load (and thus the coded caching gain),  in the proposed scheme we fix a coded caching gain $g\in[2:\Ksf^{\prime}]$ and then find the minimum needed cache size.
\begin{thm}
\label{thm:load of r>2 g general}
For an $(\Hsf,\rsf,\Msf,\Nsf)$ combination network, 
the lower convex envelop of the following points
\begin{align}
(\Msf,\Rsf)  
= 
\left(
\Nsf\frac{\binom{\Ksf^{\prime\prime}-\rsf}{q}} {\binom{\Ksf^{\prime\prime}-\rsf}{q}+\rsf\binom{\Ksf^{\prime}-1}{q-1}},
\frac{\Ksf/\Hsf (1-\Msf/\Nsf)}{g}
\right),
\label{eq:load of r>2 g general}
\end{align}
for (coded caching gain) $g\in [2:\Ksf^{\prime}]$,
$q := \Ksf^{\prime}-g+1$,  $\Ksf^{\prime} := \binom{\Hsf-1}{\rsf-1}$   and $\Ksf^{\prime\prime} := \binom{\Hsf}{\rsf-1}$, 
is achievable.
\end{thm}
\begin{thm}
\label{thm:optimaity large memory size} 
Under the same assumptions of Thm.\ref{thm:load of r>2 g general},
 when $\Msf\in\big[\frac{(\Ksf-\Hsf+\rsf-1)\Nsf}{\Ksf}, \Nsf \big]$, we have
\begin{align}
&\Rsf^{\star}=(1-\Msf/\Nsf)/\rsf.
\label{eq:optimaity large memory size}
\end{align}
\end{thm}

\begin{IEEEproof}
{\it Converse} From the general cut-set outer bound  in~\cite[Thm.2]{cachingincom} with $\alpha=1$ and $l=1$, we have $\Rsf^{\star}\geq (1-\Msf/\Nsf)/\rsf$. Between $\Msf\in \big[\frac{(\Ksf-\Hsf+\rsf-1)\Nsf}{\Ksf}, \Nsf \big]$, the outer bound is a straight line. When $\Msf=\frac{(\Ksf-\Hsf+\rsf-1)\Nsf}{\Ksf},$ we have 
\begin{align}
\Rsf^{\star}
\geq \frac{1-(\Ksf-\Hsf+\rsf-1)/\Ksf}{\rsf}
=\frac{\Hsf-\rsf+1}{\rsf\Ksf}.
\label{eq:converse}
\end{align}
{\it Achievability}
When $q=1$ in Thm.\ref{thm:load of r>2 g general}, we have 
$\Msf
=\frac{\Ksf^{\prime\prime}-r}{\Ksf^{\prime\prime}}
=\frac{(\Ksf-\Hsf+\rsf-1)\Nsf}{\Ksf}$ and
$g
=\Ksf^{\prime}
=\frac{\Ksf\rsf}{\Hsf}$. Thus from~\eqref{eq:load of r>2 g general} 
we have 
\begin{align}
&\Rsf^{\star}
\leq  \frac{\Ksf(1-\Msf/\Nsf)}{\Hsf g}
= \frac{\Hsf-\rsf+1}{\rsf\Ksf},
\label{eq:achievability}
\end{align}
which coincides with~\eqref{eq:converse}. 
The memory sharing is then used between $\Msf=\frac{(\Ksf-\Hsf+\rsf-1)\Nsf}{\Ksf}$ and $\Msf=\Nsf$.
\end{IEEEproof}

\begin{rem}
\label{rem:comparison to Yener paper} 
For the scheme in~\cite{Zewail2017codedcaching}, in order to achieve a coded caching gain $g\in [2:\Ksf^{\prime}]$ the minimum needed cache size is $\Msf=\frac{\Hsf(g-1)\Nsf}{\rsf\Ksf}$. 
When $\Ksf^{\prime}-\left\lfloor \frac{\Hsf}{\rsf-1}\right\rfloor +1\leq g \leq\Ksf^{\prime}$, the minimum cache size 
for our scheme in Thm.\ref{thm:load of r>2 g general} is strictly less than the one in~\cite{Zewail2017codedcaching};
moreover when $\rsf=2$, we have $\Ksf^{\prime}-\frac{\Hsf}{\rsf-1}+1=0$, and  for any memory size the proposed scheme is  better than~\cite{Zewail2017codedcaching}. 
 The  proof can be found in the extended version of this paper.
\end{rem}

\section{Proof of Thm.\ref{thm:load of r>2 g general}}
\label{sec:proof of thm 1}
\paragraph*{Uncoded cache placement}
If each user directly store some bits of files in the cache, the placement is uncoded. 
When placement is uncoded, each file can be effectively partitioned as
$F_i = \{F_{i,\Wc}:\Wc\subseteq [\Ksf]\}$ where $F_{i,W}$   represents the bits of $F_{i} $ which are only cached by users in $\Wc$.
\paragraph*{MAN placement}
For $\Msf=t\frac{\Nsf}{\Ksf}$ where $t\in[0:\Ksf]$, 
each file $F_i$ where $i\in[\Nsf]$ is divided into $\binom{\Ksf}{t}$ non-overlapping subfiles
of length $\frac{\Bsf}{\binom{\Ksf}{t}}$ bits; $F_i=\{F_{i,\Wc}:\Wc\subseteq[\Ksf], |\Wc|=t\}$.

 With MAN placement, a delivery scheme to create multicast messages by leveraging the symmetries in the topology 
was proposed in~\cite{wan2017novelmulticase}; 
in Section~\ref{sub:SRDS}, we revisit it. 
In Section~\ref{sub:max g}, we describe the proposed scheme to achieve coded caching gain $g=\Ksf^{\prime}$. 
In Section~\ref{sub:any g}, we generalize the scheme to any coded caching gain $g\in [2:\Ksf^{\prime}]$. 

\subsection{Separate Relay Decoding delivery Scheme (SRDS)~\cite{wan2017novelmulticase}}
\label{sub:SRDS}
In the delivery phase, user $k\in[\Ksf]$ should recover $F_{d_k,\Wc}$ for all $\Wc\subseteq [\Ksf]\setminus \{k\}$. For each such subfile $F_{d_k,\Wc}$, we find  $\Sc_{k,\Wc} :=\max_{h\in \Hc_k}|\Uc_h \cap \Wc|$ (i.e., the set of relays $\Sc_{k,\Wc}\subseteq \Hc_k$ each relay in which is connected to the largest number of users in $\Wc$). We  partition $F_{d_{k},\Wc}$ into $|\Sc_{k,\Wc}|$ equal-length pieces and denote $F_{d_{k},\Wc}=(F^{|\Sc_{k,\Wc}|}_{d_{k},\Wc,h} : h\in \Sc_{k,\Wc})$. For each relay $h\in \Sc_{k,\Wc}$, we add $F^{|\Sc_{k,\Wc}|}_{d_{k},\Wc, h}$ to $\Tc^{h}_{k,\Wc\cap\Uc_h}$; here $\Tc^{h}_{k,\Wc\cap\Uc_h}$ represents the set of bits needed to be recovered by user $k$ (first entry in the subscript) from relay $h$ (superscript) and already known by the users in $\Wc\cap\Uc_h$ (second entry in the subscript) who are also connected to relay $h$ (superscript).

The next step is to generate multicast messages. For each relay $h\in[\Hsf]$ and each set $\Jc\subseteq \Uc_h$, the server forms the multicast messages
\begin{align}
W_{\Jc}^{h} := \underset{k\in\Jc}{\oplus}\Tc^{h}_{k,\Jc\setminus \{k\}},
\label{eq:multicast messages}
\end{align} 
 where we used the same convention as that in the literature  when it comes to `summing' sets.
The message $W_{\Jc}^{h}$ is sent to relay $h$, which then forwards it to the users in $\Jc$.

The main limitation of SRDS with MAN placement is that the delivery of some subfiles, due to the network topology, needs more bits than than others--see  Example~\ref{ex:example for any g} later on.
In the next subsection, we propose a novel placement so that all multicast messages need the same amount of transmitted bits to be delivered to the intended users, in other words, all multicast messages have the same ``multicasting opportunities'' and thus the delivey phase is ``symmetric''.

\subsection{Novel Caching Scheme for $g=\Ksf^{\prime}$} 
\label{sub:max g}
We start by describing by way of an example  to achieve the maximal coded caching gain $g=\Ksf^{\prime}$;
in this case, the coded multicast gain $g$ equals the number of users connected to a relay. 

\begin{example}[$\Hsf=5$, $\rsf=3$, $\Nsf=10$, $g=6$]
\label{ex:example 1 g max}
In this example, we have $\Nsf=\Ksf=10$ and 
\begin{align*}
  &\Uc_{1}=[6], \ \Uc_{2}=\{1, 2, 3, 7, 8, 9\},\ \Uc_{3}=\{1, 4, 5, 7, 8, 10\},
\\&\Uc_{4}=\{2, 4, 6, 7, 9, 10\},\ \Uc_{5}=\{3, 5, 6, 8, 9, 10\}.
\end{align*}
We aim to achieve coded caching gain $g=6$, that is, every multicast message is simultaneously useful for $g=6$ users.
Since $\rsf=3$ (each user is connected to three relays), we can see that every $\rsf-1=2$ relays (denoted by $\Yc$) have $\binom{\Hsf-(\rsf-1)}{\rsf-(\rsf-1)}=\Hsf-\rsf+1=3$ common connected users (denoted by $\Pc_{\Yc}$). Besides the relays in $\Yc$, each of these three users is connected to a different relay other than the two relays in $\Yc$. For one user $k\in \Yc$, we assume user $k$ is connected to relay $h$ where $h\in \Hc_k \setminus \Yc$; since relay $h$ is connected to only one user (user $k$) in $\Pc_{\Yc}$,
we have $([\Ksf]\setminus \Pc_{\Yc})\cap \Uc_{h}=\Uc_{h} \setminus \{k\}$. 
\footnote{
For example, if $\Yc=\{1,2\}$, we have $\Pc_{\{1,2\}}=\{1,2,3\}$;
let us focus on user $1$, who is also connected to relay $3$;
we can see $([\Ksf]\setminus \Pc_{\{1,2\}})\cap \Uc_3=\{4,5,7,8,10\}=\Uc_{3}\setminus \{1\}$. 
}
This motivates the following placement, which considers all the $\Ksf^{\prime\prime}:=\binom{\Hsf}{\rsf-1}$ subsets  of relays with cardinality $\rsf-1$.

\paragraph*{Placement phase}
We divide each $F_{i}$ into $\Ksf^{\prime\prime}=10$ non-overlapping and equal-length pieces and denote
\begin{align*}
&  F_{i} = (F_{i,[\Ksf]\setminus\Pc_{\Yc}}:\Yc \subseteq [\Hsf], |\Yc|=\rsf-1)
\\&=\big\{F_{i,[10]\setminus\{1,2,3\}},F_{i,[10]\setminus\{1,4,5\}},F_{i,[10]\setminus\{2,4,6\}},F_{i,[10]\setminus\{3,5,6\}},
\\&\qquad F_{i,[10]\setminus\{1,7,8\}},F_{i,[10]\setminus\{2,7,9\}},F_{i,[10]\setminus\{3,8,9\}}, F_{i,[10]\setminus\{4,7,10\}},
\\&\qquad F_{i,[10]\setminus\{5,8,10\}},F_{i,[10]\setminus\{6,9,10\}}\big\}. 
\end{align*}  
It can be seen that the required memory size  is $\Msf=\Nsf\big(1-\binom{\rsf}{\rsf-1}/\Ksf^{\prime\prime} \big)=7$.  
Note that, compared to MAN placement, not all subfiles $F_{i,\Wc}$ where $|\Wc|=7$ are present.

\paragraph*{Delivery phase}
Assume $\mathbf{d}=(1:10)$. We use SRDS to let each user $k$ recover $F_{d_k, \Wc}$ where $k\notin \Wc$. 
For example, user $1$ should recover $F_{1,[10]\setminus\{1,2,3\}}$, $F_{1,[10]\setminus\{1,4,5\}}$ and $F_{1,[10]\setminus\{1,7,8\}}$. For $F_{1,[10]\setminus\{1,2,3\}}=F_{1,[4:10]}$,  we can see that relay~$1$ is  connected to users $[1:6] \cap [4:10] = \{4,5\}$,  
relay~$2$ is  connected to users $\{1,2,3,7,8,9\} \cap [4:10] = \{7,8,9\}$ while relay~$3$ is  connected to users $\{1, 4, 5, 7, 8, 10\} \cap [4:10] = \{4,5,7,8,10\}$ and 
thus we have $\Sc_{1,[4:10]}=\arg\max_{h\in \Hc_1}|\Uc_h \cap [4:10]|=\{3\}$. 
Therefore
 $F_{1,[4:10]} \in \Tc^{3}_{1,\{4,5,7,8,10\}}$. Similarly, 
 $F_{1,[10]\setminus\{1,4,5\}} \in \Tc^{2}_{1,\{2,3,7,8,9\}}$ and 
 $F_{1,[10]\setminus\{1,7,8\}} \in \Tc^{1}_{1,\{2,3,4,5,6\}}$. 
After considering all the subfiles demanded by all  users, for relay $h=1$ (and similarly for all other relays) we have
\begin{align*}
&\Tc^{1}_{1,\{2,3,4,5,6\}}\negmedspace=\negmedspace\big\{\negmedspace F_{1,[10]\setminus\{1,7,8\}} \negmedspace\big\},  \negmedspace \Tc^{1}_{2,\{1,3,4,5,6\}}\negmedspace=\negmedspace\big\{\negmedspace F_{2,[10]\setminus\{2,7,9\}} \negmedspace\big\},\\
&\Tc^{1}_{3,\{1,2,4,5,6\}}\negmedspace=\negmedspace\big\{\negmedspace F_{3,[10]\setminus\{3,8,9\}} \negmedspace\big\},  \negmedspace \Tc^{1}_{4,\{1,2,3,5,6\}}\negmedspace=\negmedspace\big\{\negmedspace F_{4,[10]\setminus\{4,7,10\}} \negmedspace\big\},\\
&\Tc^{1}_{5,\{1,2,3,4,6\}}\negmedspace\negmedspace=\negmedspace\big\{\negmedspace F_{5,[10]\setminus\{5,8,10\}} \negmedspace\big\},  \negmedspace \Tc^{1}_{6,\{1,2,3,4,5\}}\negmedspace\negmedspace=\negmedspace\big\{\negmedspace F_{6,[10]\setminus\{6,9,10\}} \negmedspace\big\}.
\end{align*}

For each relay $h\in[\Hsf]$, we create the multicast messages $W_{\Uc_h}^{h}$ as in~\eqref{eq:multicast messages}
to be sent to relay $h$ and then forwarded to the users in $\Uc_h$.
Notice that in this example, by the novel placement, each subfile is multicasted with other $5$ subfiles and thus the coded caching gain is $g=6$. 
The achieved max-link load is $1/10$, which coincides with the cut-set outer bound in~\cite{cachingincom};
the max-link load in~\cite{Zewail2017codedcaching} is $0.118$.  In this example, the proposed placement is uncoded and is information theoretically optimal.
\end{example}

We now generalize the scheme in Example~\ref{ex:example 1 g max} to achieve the maximal coded caching gain $g=\Ksf^{\prime}$. 
\paragraph*{Placement phase}
Each file $F_i$ is divided into $\Ksf^{\prime\prime}:=\binom{\Hsf}{\rsf-1}$ non-overlapping and equal-length pieces denoted by $F_{i}=(F_{i,[\Ksf]\setminus\Pc_{\Yc}}:\Yc \subseteq [\Hsf], |\Yc|=\rsf-1)$. 
$F_{i,\Wc}, i\in[\Nsf],$ is cached by user $k$ if $k\in \Wc$, which requires
$\Msf=\Nsf\big(1-\binom{\rsf}{\rsf-1}/\Ksf^{\prime\prime} \big)=\frac{(\Ksf-\Hsf+\rsf-1)\Nsf}{\Ksf}$ 
(since  $|\Pc_{\Yc}|=\binom{\Hsf-(\rsf-1)}{\rsf-(\rsf-1)}=\Hsf-\rsf+1$). 

\paragraph*{Delivery phase}
User $k$ should recover $F_{d_k,[\Ksf]\setminus\Pc_{\Yc}}$ where  $\Yc \subseteq [\Hsf]$ with $|\Yc|=\rsf-1$ and $k\in \Pc_{\Yc}$. It can be seen that if and only if $\Yc\subseteq \Hc_k$, we have $k\in \Pc_{\Yc}$. So we need to consider each user $k\in[\Ksf]$ and each set of relays $\Yc \subseteq \Hc_k$ with cardinality $|\Yc|=\rsf-1$. We can see that $|\Hc_k \setminus \Yc|=1$ and let $h\in \Hc_k \setminus \Yc$.  Besides $\Yc$, each user in $\Pc_{\Yc}$ is connected to a different relay other than the relays in $\Yc$. Hence, $\Pc_{\Yc} \cap \Uc_{h}=\{k\}$ and thus $([\Ksf]\setminus \Pc_{\Yc})\cap \Uc_{h}=\Uc_{h} \setminus \{k\}$. Each relay  $h^{\prime}\in (\Hc_k\setminus\{h\})$ is connected to $|\Pc_{\Yc}|=\Hsf-\rsf+1$ users in $\Pc_{\Yc}$ and thus $|([\Ksf]\setminus \Pc_{\Yc})\cap \Uc_{h^{\prime}}|=|\Uc_{h^{\prime}}|-|\Pc_{\Yc}|$. So for $F_{d_k,[\Ksf]\setminus \Pc_{\Yc}}$, we have $\Sc_{k,[\Ksf]\setminus \Pc_{\Yc}}=\{h\}$ and put it 
 in $\Tc^{h}_{k,\Uc_h \setminus \{k\}}$.  

For each relay $h\in[\Hsf]$, the server forms the multicast messages
as in~\eqref{eq:multicast messages}
and transmits it to relay $h$, which then forwards it to  users in $\Uc_h$.

\paragraph*{Max-link load} 
Each demanded subfile is multicasted with other $\Ksf^{\prime}-1$ subfiles and thus $g=\Ksf^{\prime}$. As a result, the max-link load is as in~\eqref{eq:load of r>2 g general}.

\subsection{Generalization to $g\in [2:\Ksf^{\prime}]$} 
\label{sub:any g}
We now extend the scheme in Section~\ref{sub:max g} to any $g\in [2:\Ksf^{\prime} ]$.
The novel ingredient here is an additional `precoding' of the files before placement, i.e., in other words, the design of a coded placement  based on the topology of the network instead of uncoded placement. We start with an example.
%

\begin{example}[$\Hsf=4$, $\rsf=2$, $\Nsf=6$, $g=2$] 
\label{ex:example for any g}
In Example~\ref{ex:example 1 g max}, for each collection $\Qc$ of  $q$ subsets of relays with cardinality $\rsf-1$ each, we have one corresponding subfile. 
Similar to Example~\ref{ex:example 1 g max}, for each set of $\rsf-1=1$ relay, in this example we also determine the set of common connected users, in this case $\Pc_{\{h\}}=\Uc_h$, i.e., $\Pc_{\{1\}}=\{1,2,3\}$, $\Pc_{\{2\}}=\{1,4,5\}$, $\Pc_{\{3\}}=\{2,4,6\}$, and $\Pc_{\{4\}}=\{3,5,6\}$. In addition, we have $q=\Ksf^{\prime}-g+1=2$. 
Before introducing the additional MDS precoding, we show  that, if we proceed as for the previous example, not all the subfiles are sent in a linear combination involving the same number of subfiles, in other words, not all subfiles have the same ``multicasting opportunities.''

Consider  $q=2$ subsets of relays each with cardinality $\rsf-1=1$, e.g., $\Yc_1=\{1\}$ and $\Yc_2=\{3\}$. Since besides relay $1$, each user in $\Pc_{\Yc_1}$ is connected to a different relay other than relay $1$, we have $|\Pc_{\Yc_1}\cap \Uc_2|=1$. Similarly, we have $|\Pc_{\Yc_2}\cap \Uc_2|=1$. It can also be checked that $\Pc_{\Yc_1}\cap \Uc_2\neq \Pc_{\Yc_2}\cap \Uc_2$. Hence, we have $|(\Pc_{\Yc_1}\cup \Pc_{\Yc_2})\cap \Uc_2|=q=2$ and thus $|\big([\Ksf]\setminus (\Pc_{\Yc_1}\cup \Pc_{\Yc_2})  \big) \cap \Uc_2|=|\Uc_2|-q=1$. 
Hence with SRDS, for user $1$ we can transmit $F_{d_1,[\Ksf]\setminus (\Pc_{\Yc_1}\cup \Pc_{\Yc_2})}=F_{d_1,\{5\}}$ and $F_{d_5,\{1\}}$ simultaneously in one linear combination. Similarly, it can be seen that the subfiles $F_{d_1,[\Ksf]\setminus (\Pc_{\{1\}}\cup \Pc_{\{4\}})}=F_{d_1,\{4\}}$, $F_{d_1,[\Ksf]\setminus (\Pc_{\{2\}}\cup \Pc_{\{3\}})}=F_{d_1,\{3\}}$ and $F_{d_1,[\Ksf]\setminus (\Pc_{\{2\}}\cup \Pc_{\{4\}})}=F_{d_1,\{2\}}$ demanded by user $1$ have the same ``multicasting opportunities'' as $F_{d_1,\{4\}}$.

However, consider the following $q=2$ subsets of relays each with cardinality $\rsf-1=1$: 
$\Yc_3=\{1\}$ and $\Yc_4=\{2\}$. For user $1$ who does not know $F_{d_1,[\Ksf]\setminus (\Pc_{\Yc_3}\cup \Pc_{\Yc_4})}=F_{d_1,\{6\}}$, since $\Hc_1\subseteq \Yc_3\cup \Yc_4$, we can see that for each $h\in \Hc_1$ we have  $|(\Pc_{\Yc_3}\cup \Pc_{\Yc_4})\cap \Uc_h|=3$ and so $|\big([\Ksf]\setminus (\Pc_{\Yc_3}\cup \Pc_{\Yc_4})  \big) \cap \Uc_h|=0$. In other words, with SRDS to transmit $F_{d_1,\{6\}}$, we cannot transmit other subfiles in the same combination.   
 
Hence, the main idea of our proposed scheme is to let user $1$ recover $F_{d_1,\{2\}}$, $F_{d_1,\{3\}}$, $F_{d_1,\{4\}}$ and $F_{d_1,\{4\}}$ in the delivery phase, and ignore $F_{d_1,\{6\}}$ which has less ``multicasting opportunities''.
Notice that the subfile $F_{d_1,[\Ksf]\setminus (\Pc_{\{3\}}\cup \Pc_{\{4\})}}=F_{d_1,\{1\}}$ is cached by user $1$.
This motivates the following placement.

\paragraph*{Placement phase}
Each file $F_i$ is divided into $1+4=5$ non-overlapping and equal-length pieces, which are then encoded by using a $(6,5)$ MDS code (not the $(\Hsf,\rsf)=(4,2)$ MDS code as in~\cite{Zewail2017codedcaching}). Each MDS coded symbol of $F_{i}$ is cached by one user $k\in[\Ksf]$ and is denoted by $f_{i,\{k\}}$, which contains $\Bsf/5$ bits. So the cache size needs to be $\Msf=6/5$. 

\paragraph*{Delivery phase}
Assume $\mathbf{d}=(1:6)$. We use SRDS to let each user $k\in[\Ksf]$ recover $f_{d_k,[\Ksf]\setminus(\Pc_{\Yc_1}\cup \Pc_{\Yc_2})}$ where $k\notin [\Ksf]\setminus (\Pc_{\Yc_1}\cup \Pc_{\Yc_2})$ and $\Hc_k \nsubseteq (\Yc_1\cup \Yc_2)$,
such that from placement and delivery phases, each user can obtain $5$ MDS coded symbols of file $F_{d_k}$ and is thus able to recover $F_{d_k}$. 
For example, user $1$ must recover $f_{1,\{2\}}$, $f_{1,\{3\}}$,  $f_{1,\{4\}}$, and $f_{1,\{5\}}$; those, together with the cached MDS coded symbol $f_{1,\{1\}}$, allows him to recover $F_{1}$.
For $f_{1,\{2\}}$, we can see that relay~1 is  connected to user $\{2\} \cap \{1,2,3\} = \{2\}$, while 
relay~2 is  connected to user $\{2\} \cap \{1,4,5\} = \emptyset$, and 
thus we have $\Sc_{1,\{2\}}=\arg\max_{h\in \Hc_1}|\Uc_h \cap \{2\}|=\{1\}$ and $f_{1,\{2\}}\in\Tc^{1}_{1,\{2\}}$. 
After considering all the subfiles demanded by all the users, for relay $h=1$ (and similarly for all other relays) we have
\begin{align*}
 &\Tc^{1}_{1,\{2\}} =\{f_{1,\{2\}}\}, \  \Tc^{1}_{1,\{3\}}  =\{f_{1,\{3\}}\}, \ \Tc^{1}_{2,\{1\}} =\{f_{2,\{1\}}\},
 \\
 & \Tc^{1}_{2,\{3\}}  =\{f_{2,\{3\}}\}, \ \Tc^{1}_{3,\{1\}} =\{f_{3,\{1\}}\}, \  \Tc^{1}_{3,\{2\}}  =\{f_{3,\{2\}}\}.
\end{align*}

We then create the multicast messages as in~\eqref{eq:multicast messages} for each $\Jc\subseteq \Uc_h$ where $|\Jc|=g=2$. For example, the server transmits to relay~1
 $W_{\{1,2\}}^{1}=\Tc^{1}_{1,\{2\}}\oplus \Tc^{1}_{2,\{1\}}= f_{1,\{2\}}\oplus f_{2,\{1\}},\ W_{\{1,3\}}^{1}=\Tc^{1}_{1,\{3\}}\oplus \Tc^{1}_{3,\{1\}}=f_{1,\{3\}}\oplus f_{3,\{1\}},\ W_{\{2,3\}}^{1}=\Tc^{1}_{2,\{3\}}\oplus \Tc^{1}_{3,\{2\}}=f_{2,\{3\}}\oplus f_{3,\{2\}}$, which are then forwarded to the demanding users. 
The achieved max-link load is $3/5$, while that of~\cite{Zewail2017codedcaching} is $9/10$.  
The outer bound idea used in~\cite{cachingincom}, which leverages the cut-set bound from~\cite{dvbt2fundamental}, can be straightforwardly extended to leverage  the tighter outer bound from~\cite{yas2}; by doing so, for this example we obtain as outer bound $3/5$;
therefore, our proposed scheme is optimal. 
In this example, thanks to the novel placement, each MDS coded symbol is multicasted with another one and thus the coded caching gain is $g=2$.

Notice that the outer bound under the constraint of uncoded placement in~\cite[Thm.4]{novelwan2017}  is $157/255\approx 0.616$, that is, in this example using uncoded cache placement is strictly suboptimal.
\end{example}

We now present our novel scheme that attains  $g\in [2:\Ksf^{\prime}]$. 

\paragraph*{Placement phase}
Recall $q=\Ksf^\prime-g+1$, $\Ksf^\prime=\binom{\Hsf-1}{\rsf-1}$ and $\Ksf^{\prime\prime} := \binom{\Hsf}{\rsf-1}$.
Each file $F_i, i\in[\Nsf],$ is divided into $\binom{\Ksf^{\prime\prime}-\rsf}{q}+\rsf\binom{\Ksf^\prime-1}{q-1}$ non-overlapping and equal-length pieces, which are then encoded by using a $\Big(\binom{\Ksf^{\prime\prime}}{q},\binom{\Ksf^{\prime\prime}-\rsf}{q}+\rsf\binom{\Ksf^\prime-1}{q-1}\Big)$ MDS code. 
For each collection $\Qc$ including $q$ subsets of relays with cardinality $\rsf-1$ each, there is an MDS coded symbol $f_{i,[\Ksf]\setminus \underset{\Yc\in \Qc}{\cup}\Pc_{\Yc}}$ cached by users in $[\Ksf]\setminus \underset{\Yc\in \Qc}{\cup}\Pc_{\Yc}$. The required memory size to store these MDS coded symbols is $\Msf=\Nsf\frac{\binom{\Ksf^{\prime\prime}-\rsf}{q}}{\binom{\Ksf^{\prime\prime}-\rsf}{q}+\rsf\binom{\Ksf^\prime-1}{q-1}}$. 
After the placement phase, since there are $\binom{\Ksf^{\prime\prime}}{q}$ collections of $q$ subsets of relays each of cardinality $\rsf-1$, each file $F_i$ has $\binom{\Ksf^{\prime\prime}}{q}$ MDS coded symbols.
For each user $k$, we divide  the collections $\Qc$ into $3$ classes.
\begin{enumerate}
\item Class 1: if there is no $\Yc \in \Qc$ such that $\Yc \subseteq \Hc_k$, we have $k\in[\Ksf]\setminus \underset{\Yc\in \Qc}{\cup}\Pc_{\Yc}$ and thus the symbol $f_{i,[\Ksf]\setminus \underset{\Yc\in \Qc}{\cup}\Pc_{\Yc}}$ is cached by user $k$ where $i\in[\Nsf]$. The number of collections of Class 1 is $\binom{\Ksf^{\prime\prime}-\rsf}{q}$.
\item Class 2: if $k\notin[\Ksf]\setminus \underset{\Yc\in \Qc}{\cup}\Pc_{\Yc}$ and $\Hc_k \nsubseteq \underset{\Yc\in \Qc}{\cup}\Yc$, we transmit the symbol $f_{d_k,[\Ksf]\setminus \underset{\Yc\in \Qc}{\cup}\Pc_{\Yc}}$ to user $k$ in one combination including other $g-1$ symbols
in the delivery phase. Furthermore, if and only if at least one $\Yc\in \Qc$ is a subset of $\Hc_k$, we can see that $k\notin [\Ksf]\setminus \underset{\Yc\in \Qc}{\cup}\Pc_{\Yc}$. Recall that each $\Yc\in \Qc$ has $\rsf-1 $ relays. So if $\Hc_k \nsubseteq \underset{\Yc\in \Qc}{\cup}\Yc$, at most one $\Yc\in \Qc$ is a subset of $\Hc_k$. As a result, if and only if  $\Hc_k \nsubseteq \underset{\Yc\in \Qc}{\cup}\Yc$ and there exists only one $\Yc\in \Qc$ such that $\Yc\subseteq \Hc_k$, one has
 $k\notin[\Ksf]\setminus \underset{\Yc\in \Qc}{\cup}\Pc_{\Yc}$ and $\Hc_k \nsubseteq \underset{\Yc\in \Qc}{\cup}\Yc$. Hence, the number of symbols to be recovered by user $k$ in the delivery phase is $\binom{\rsf}{\rsf-1}\binom{\binom{\Hsf-1}{\rsf-1}-1}{q-1}=\rsf\binom{\Ksf^\prime-1}{q-1}$.
\item Class 3: if $k\notin[\Ksf]\setminus \underset{\Yc\in \Qc}{\cup}\Pc_{\Yc}$ and $\Hc_k \subseteq \underset{\Yc\in \Qc}{\cup}\Yc$, for each $h\in \Hc_k$ we have $|([\Ksf]\setminus \underset{\Yc\in \Qc}{\cup}\Pc_{\Yc})\cap \Uc_h|\leq \Ksf^{\prime}-(\Hsf-\rsf+1)$. Hence, we let
user $k$ ignore the symbol $f_{d_k,[\Ksf]\setminus \underset{\Yc\in \Qc}{\cup}\Pc_{\Yc}}$.
\end{enumerate}

\paragraph*{Delivery phase}
In the delivery phase, we let each user $k\in[\Ksf]$ recover the MDS coded symbols in Class 2.
We focus on each user $k\in[\Ksf]$, each relay $h\in \Hc_k$, and each set of users $\Jc \subseteq \Uc_h$ where $|\Jc|=g$ and $k\in \Jc$.  Besides relays in $\Hc_k \setminus \{h\}$, each user in $\Pc_{\Hc_k \setminus \{h\}}$ is connected to a different relay other than the relays in $\Hc_k \setminus \{h\}$. So we have that $\Pc_{\Hc_k \setminus \{h\}}\cap \Uc_{h}=\{k\}$.
Let $\Mc:=\Uc_h\setminus \Jc$. For each user $k^{\prime}\in\Mc$, we can find a set of relays $\Hc_{k^{\prime}}\setminus \{h\}$. We can similarly prove  that $\Pc_{\Hc_{k^{\prime}}\setminus \{h\}}\cap \Uc_{h}=\{k^{\prime}\}$.  Hence, we construct the  collection
\begin{align}
\Qc^{\prime}=\big\{\Hc_{k^{\prime}}\setminus \{h\}:k^{\prime}\in \Mc\big\}\cup \big\{ \Hc_k \setminus \{h\}\big\}\label{eq:Q prime}
\end{align}
and we have  $([\Ksf]\setminus \underset{\Yc\in \Qc^{\prime}}{\cup}\Pc_{\Yc})\cap \Uc_h=\Jc\setminus\{k\}$. By this construction, we have $|\Qc^{\prime}|=\Ksf^\prime-g+1=q$. 
In addition, since $(\Hc_k \setminus \{h\})\in \Qc^{\prime}$ and there is no set in $\Qc^{\prime}$ containing relay $h$, we have that $\big\{ \Hc_k \setminus \{h\}\big\}$ is the  only  set in $\Qc^{\prime}$ which is a subset of $\Hc_k$ and that $\Hc_k \nsubseteq \underset{\Yc\in \Qc^{\prime}}{\cup}\Yc$. 
So $f_{d_k, [\Ksf]\setminus \underset{\Yc\in \Qc^{\prime}}{\cup}\Pc_{\Yc}}$ should be recovered by user $k$ and be put in $\Tc^{h}_{k,\Jc \setminus \{k\}}$. 
 As a result, for each relay $h\in \Hc_k$ and each set of users $\Jc \subseteq \Uc_h$ where $|\Jc|=g$ and $k\in \Jc$, we consider a different symbol of $F_{d_k}$ demanded by user $k$. With
$|\Uc_h|=\Ksf^\prime$ and $q=\Ksf^\prime-g+1$, we can prove that 
 in the delivery phase, we  consider all of the  $\rsf\binom{\Ksf^\prime-1}{g-1}=\rsf\binom{\Ksf^\prime-1}{\Ksf^\prime-q}=\rsf\binom{\Ksf^\prime-1}{q-1}$ symbols which are needed to be recovered by user $k$.

 For each relay $h\in[\Hsf]$ and each set $\Jc\subseteq \Uc_h$  where $|\Jc|=g$, the server forms the multicast messages $W_{\Jc}^{h}$ as in~\eqref{eq:multicast messages} and transmit it to relay $h$, who then forwards it to each user $k\in \Jc$. 

Notice that when $g=\Ksf^{\prime}$, we have $q=\Ksf^{\prime}-g+1=1$. So $\binom{\Ksf^{\prime\prime}}{q}=\binom{\Ksf^{\prime\prime}-\rsf}{q}+\rsf\binom{\Ksf^{\prime}-1}{q-1}=\Ksf^{\prime\prime}$ and thus we need not MDS precoding procedure. Hence, the above scheme is equivalent to the scheme in Section~\ref{sub:max g}.

\paragraph*{Max-link load} 
Each demanded subfile is multicasted with other $g-1$ subfiles such that the coded caching gain is $g$. As a result, the max-link load is as in~\eqref{eq:load of r>2 g general}.

\begin{rem}
\label{rem:to be improve}
From~\eqref{eq:Q prime}, since  $(\Hc_k \setminus \{h\})\in \Qc^{\prime}$,    for each $h^{\prime}\in \Hc_k \setminus \{h\}$, we have $|([\Ksf]\setminus \underset{\Yc\in \Qc^{\prime}}{\cup}\Pc_{\Yc})\cap \Uc_{h^{\prime}}| \leq\Ksf^{\prime}-(\Hsf-\rsf+1)$. Hence, 
when $g>\Ksf^{\prime}-(\Hsf-\rsf+1)+1$, we have  $|([\Ksf]\setminus \underset{\Yc\in \Qc^{\prime}}{\cup}\Pc_{\Yc})\cap \Uc_{h}|=g-1>\Ksf^{\prime}-(\Hsf-\rsf+1)$. In conclusion, $\{h\}=\arg\max_{h_1\in \Hc_k}|([\Ksf]\setminus \underset{\Yc\in \Qc^{\prime}}{\cup}\Pc_{\Yc})\cap \Uc_{h_1}|$ and the multicast message generation is identical to SRDS. However, when  $g\leq \Ksf^{\prime}-(\Hsf-\rsf+1)+1$, $h$ may not be the relay connected to the largest number of users in the considered subset and the coded caching gain may be reduced. So the future work includes the improvement for $g\leq \Ksf^{\prime}-(\Hsf-\rsf+1)+1$. 
\end{rem}

\begin{example}[$\Hsf=6$, $\rsf=2$, $\Nsf=15$]
In Fig.~\ref{fig:numerical 1}, 
we compare the performance of the proposed scheme with that of~\cite{Zewail2017codedcaching,wan2017novelmulticase,novelwan2017}  and the enhanced cut-set outer bound based on~\cite{yas2} as described in Example~\ref{ex:example for any g}. 
Notice that the proposed scheme is exactly optimal for $10\leq\Msf\leq 15$.
\begin{figure}
\centering{}
\includegraphics[scale=0.6]{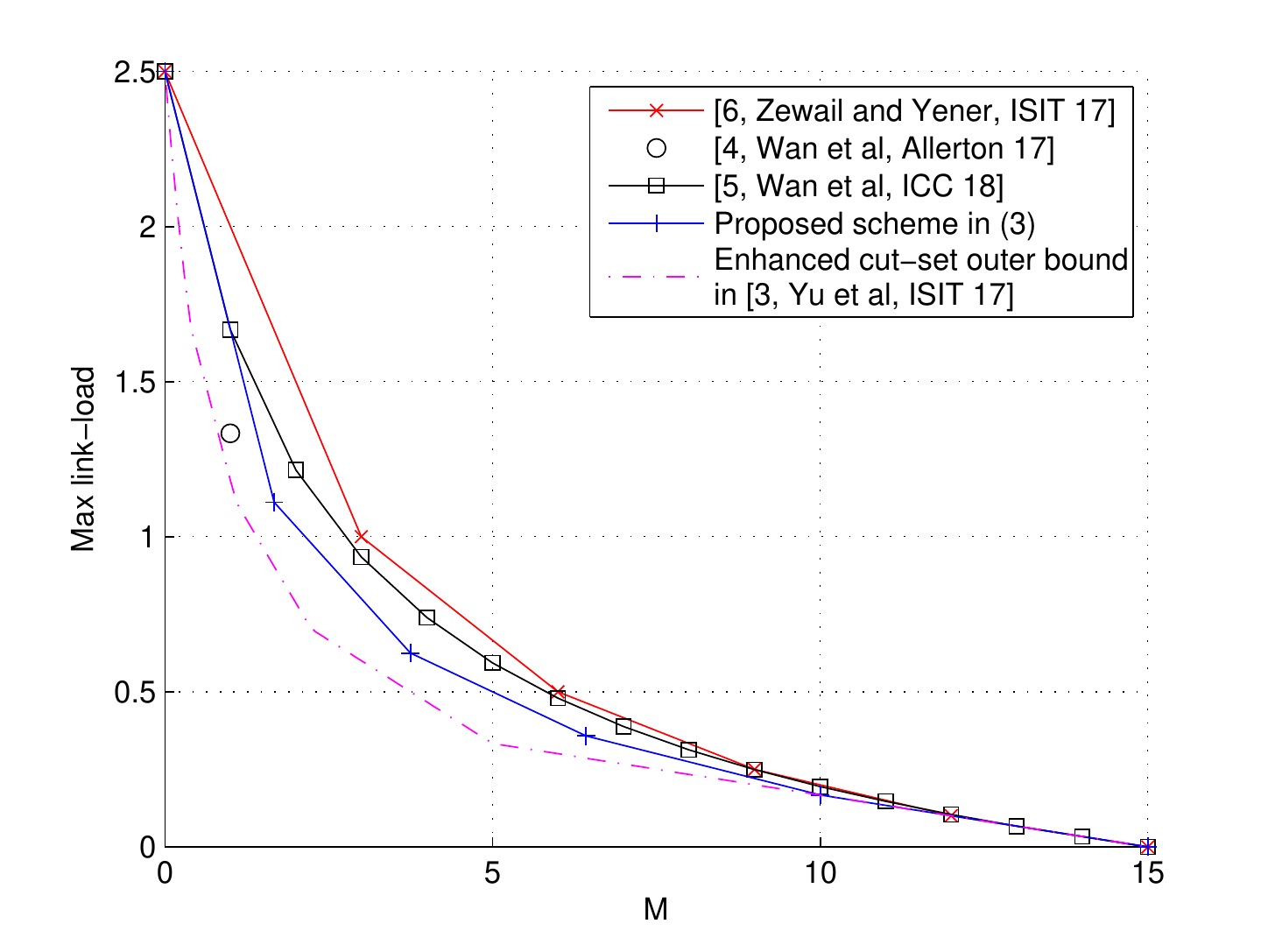}
\caption{Memory-load tradeoff for the combination network with $\Hsf=6$, $\rsf=2$ and $\Ksf=\Nsf=15$.}
\label{fig:numerical 1}
\end{figure}
\end{example}

\section{Further Improvement for Thm.\ref{thm:load of r>2 g general}}
\label{sec:improvement}
We can further improve the asymmetric coded placement proposed in Section~\ref{sub:any g}. 

It is stated in Section~\ref{sub:any g} that for each collection $\Qc$ where $|\Qc|=p$ and each element in $\Qc$ includes $\rsf-1$ relays, we can generate one MDS symbol $f_{i,[\Ksf]\setminus \underset{\Yc\in \Qc}{\cup}\Pc_{\Yc}}$. Recall that, for each user $k$, the MDS symbols can be divided into $3$ classes: the symbols which are cached by user $k$, the symbols which are needed to be recovered by user $k$ in the delivery phase and the symbols which are ignored by user $k$ in the delivery phase. Moreover, when 
$\Hc_k \subseteq \underset{\Yc\in \Qc}{\cup}\Yc$, the MDS symbol $f_{d_k,[\Ksf]\setminus \underset{\Yc\in \Qc}{\cup}\Pc_{\Yc}}$ is in Class $3$.
So if $\underset{\Yc\in \Qc}{\cup}\Yc=[\Hsf]$, there is no user who needs to recover the MDS symbol $f_{d_k,[\Ksf]\setminus \underset{\Yc\in \Qc}{\cup}\Pc_{\Yc}}$ in the delivery phase. As a result, we need not to generate the MDS symbol $f_{i,,[\Ksf]\setminus \underset{\Yc\in \Qc}{\cup}\Pc_{\Yc}}$ for the collections $\Qc$ where  $\underset{\Yc\in \Qc}{\cup}\Yc=[\Hsf]$.

In Appendix, we prove the following lemma.
\begin{lem}
\label{lem:needed memory size for improved scheme}
For each user $k\in[\Ksf]$,  the number of collections $\Qc$ where $\underset{\Yc\in \Qc}{\cup}\Yc\neq [\Hsf]$ and there is no $\Yc \in \Qc$ such that $\Yc \subseteq \Hc_k$ is
\begin{align}
&\Gsf:=\sum^{\Hsf-1}_{a=\rsf-1}(\Xsf_1+\Ysf_1+\Zsf_1)(-1)^{\Hsf-a+1},\label{eq:improved scheme cached number}\\
&\Xsf_1:=\binom{\Hsf-\rsf}{a-\rsf}\binom{\binom{a}{\rsf-1}-\rsf}{q},\label{eq:def of x1}\\
&\Ysf_1:=\rsf\binom{\Hsf-\rsf}{a-\rsf+1}\binom{\binom{a}{\rsf-1}-1}{q},\label{eq:def of y1}\\
&\Zsf_1:=\left(\binom{\Hsf}{a}-\rsf\binom{\Hsf-\rsf}{a-\rsf+1}-\binom{\Hsf-\rsf}{a-\rsf} \right)\binom{\binom{a}{\rsf-1}}{q}.\label{eq: def of z1}
\end{align} 
\end{lem}

It can be computed that the number of collections $\Qc$ where  $\Hc_k \subseteq \underset{\Yc\in \Qc}{\cup}\Yc$ and there is only one set $\Yc \in\Qc$ where $\Yc\subseteq \Hc_k$ is also $\rsf\binom{\Ksf^{\prime}-1}{q-1}$. For each of such collections $\Qc$, user $k$ needs to recover the MDS symbol $f_{d_k,[\Ksf]\setminus \underset{\Yc\in \Qc}{\cup}\Pc_{\Yc}}$. Hence, we can derive the following theorem.

\begin{thm}
\label{thm:improved scheme general}
For an $(\Hsf,\rsf,\Msf,\Nsf)$ combination network, 
the lower convex envelop of the following points
\begin{align}
(\Msf,\Rsf)  
= 
\left(
\Nsf\frac{\Gsf} {\Gsf+\rsf\binom{\Ksf^{\prime}-1}{q-1}},
\frac{\Ksf/\Hsf (1-\Msf/\Nsf)}{g}
\right),
\label{eq:improved scheme general}
\end{align}
for (coded caching gain) $g\in [2:\Ksf^{\prime}]$,
$q := \Ksf^{\prime}-g+1$ and  $\Ksf^{\prime} := \binom{\Hsf-1}{\rsf-1}$   
is achievable.
\end{thm}

\appendices
\section{Proof of Remark~\ref{rem:comparison to Yener paper}}
\label{sec:proof of remark}
We give the outline of the proof.
We aim to prove that for a coded caching gain $g\geq \Ksf^{\prime}-\left\lfloor \frac{\Hsf}{\rsf-1}\right\rfloor+1$ meaning that $q\leq \left\lfloor \frac{\Hsf}{\rsf-1}\right\rfloor$, the needed minimum memory size of the proposed scheme is less than the one in~\cite{Zewail2017codedcaching}, i.e.,
\begin{align}
\Nsf\frac{\binom{\Ksf^{\prime\prime}-\rsf}{q}} {\binom{\Ksf^{\prime\prime}-\rsf}{q}+\rsf\binom{\Ksf^{\prime}-1}{q-1}}< \frac{\Hsf(g-1)\Nsf}{\rsf\Ksf}.\label{eq:comparison to Zewail}
\end{align}
It is equivalent to prove 
\begin{align}
1-\frac{\rsf\binom{\Ksf^{\prime}-1}{q-1}} {\binom{\Ksf^{\prime\prime}-\rsf}{q}+\rsf\binom{\Ksf^{\prime}-1}{q-1}}&< \Big(\Ksf^{\prime}-q\Big)\frac{\Hsf}{\rsf\Ksf}\nonumber\\
&=(\Ksf-\Hsf q/\rsf)/\Ksf=1-\Hsf q/\rsf\Ksf. \label{eq:equivalent to prove 1}
\end{align}
It is equivalent to prove
\begin{align}
&\frac{1}{x/(q\rsf)+1}-\frac{\Hsf q}{\rsf \Ksf}>0,\label{eq:equivalent to prove 2}\\
&\textrm{where } x=\frac{\big(\Ksf^{\prime\prime}-r\big)\times\dots\times \big(\Ksf^{\prime\prime}-r-q+1\big)}{\big(\Ksf^{\prime}-1\big)\times\dots\times \big(\Ksf^{\prime}-q+1\big)}.\nonumber
\end{align}
It is equivalent to prove
\begin{align}
x-\frac{\Ksf\rsf^{2}}{\Hsf}+q\rsf=x-\Ksf^{\prime}\rsf+q\rsf<0.\label{eq:equivalent to prove 3}
\end{align}
Since $\frac{\Hsf}{\Hsf-\rsf+1}<\frac{\rsf+q-1}{q-1}$, we have 
\begin{align*}
x&<\Big(\Ksf^{\prime\prime}-\rsf \big)\big(\frac{\Hsf}{\Hsf-\rsf+1}\Big)^{q-1}\\
&<\Ksf^{\prime\prime}\big(\frac{\Hsf}{\Hsf-\rsf+1}\big)^{q-1}\\
&=\Ksf^{\prime}\big(\frac{\Hsf}{\Hsf-\rsf+1}\big)^{q}.
\end{align*}
Hence, it is sufficient to prove 
\begin{align*}
\Ksf^{\prime}\big(\frac{\Hsf}{\Hsf-\rsf+1}\big)^{q}-\Ksf^{\prime}\rsf+q\rsf<0.
\end{align*}
Since $\frac{q\rsf}{\Ksf^{\prime}}<1$, it is sufficient to prove
\begin{align*}
\big(\frac{\Hsf}{\Hsf-\rsf+1}\big)^{q}-\rsf+1<0.
\end{align*}
Since $q\leq \left\lfloor \frac{\Hsf}{\rsf-1}\right\rfloor$, it is sufficient to prove
\begin{align*}
\big(\frac{\Hsf}{\Hsf-\rsf+1}\big)^{\Hsf/(\rsf-1)}-(\rsf-1)<0.
\end{align*}
Let $\rsf-1=u$. We can see that $\Hsf\geq u+2$. It is equivalent to prove
\begin{align*}
\big(\frac{\Hsf}{\Hsf-u}\big)^{\Hsf/u}-u<0.
\end{align*}
Let $G(\Hsf,u)=\big(\frac{\Hsf}{\Hsf-u}\big)^{\Hsf/u}-u$. We then compute the partial derive of $G(\Hsf,u)$ with respect to $\Hsf$ denoted by $\frac{\partial G(\Hsf,u)}{\partial \Hsf}$,
\begin{align*}
\frac{\partial G(\Hsf,u)}{\partial \Hsf}=\big(\frac{\Hsf}{\Hsf-u}\big)^{\Hsf/u}\frac{1}{u(\Hsf-u)}\big(-u+(\Hsf-u)\log (\frac{\Hsf}{\Hsf-u}) \big).
\end{align*}
 By Taylor series, we can prove that 
 \begin{align*}
 e^y>\big(\frac{\Hsf}{\Hsf-u}\big)^{\Hsf-u},
 \end{align*}
 where $e$ is the Euler's number.
Hence $\frac{\partial G(\Hsf,u)}{\partial \Hsf}<0$ and thus $G(\Hsf,u)$ is monotonically decreasing with respect to $\Hsf$. Similarly by the partial derive of $G(\Hsf,u)$ with respect to $u$, we can see that for any $\Hsf^{\prime}$ and $u^{\prime}$, if $G(\Hsf^{\prime},u^{\prime})<0$, we have $G(\Hsf^{\prime},u)<G(\Hsf^{\prime},u^{\prime})$ where $u>u^{\prime}$. From these two monotonic results, we can have
\begin{enumerate}
\item When $u=7$, we have $G(u+2,u)<0$. So for any $u\geq 7$  and any $\Hsf/geq u+2$, we have $G(\Hsf,u)<0$. So we can prove the proposed scheme is strictly better than~\cite{Zewail2017codedcaching}.
\item When $u<7$, we can find the minimum value $v_u$ where $G(u+v_u,u)<0$. Thus for any $\Hsf \geq u+v_u$,  $G(\Hsf,u)<0$ and we can prove our results. For $\Hsf <u+v_u$, we can directly compute and show~\eqref{eq:comparison to Zewail} is true. 
\end{enumerate} 
In conclusion, we prove~\eqref{eq:comparison to Zewail}.

\section{Proof of Lemma~\ref{lem:needed memory size for improved scheme}}
We focus on one integer $a\in [\rsf-1,\Hsf-1]$ and one user $k\in[\Ksf]$. We can divide  collections $\Qc$ where $|\underset{\Yc\in \Qc}{\cup}\Yc|=a$ and there is no $\Yc \in \Qc$ such that $\Yc \subseteq \Hc_k$ into $3$ groups:
\begin{enumerate}
\item $|\Hc_k\cap \underset{\Yc\in \Qc}{\cup}\Yc|=\rsf$. The number of set $\Sc\subseteq [\Hsf]$ where $|\Sc|=a$ and $\Hc_k \subseteq \Sc$  is $\binom{\Hsf-\rsf}{a-\rsf}$. For each of thus sets $\Sc$, we want to choose one collection $\Qc$ 
where $|\Qc|=q$ and for each $\Yc\in \Qc$ we have $\Yc\subseteq \Sc$ and $\Yc\nsubseteq \Hc_{k}$.
It can be computed that the number of such collections for each set $\Sc$ is $\binom{\binom{a}{\rsf-1}-\rsf}{q}$. Hence, the number of collections $\Qc$ in this group is $\Xsf_1$.

\item $|\Hc_k\cap \underset{\Yc\in \Qc}{\cup}\Yc|=\rsf-1$. The number of set $\Sc\subseteq [\Hsf]$ where $|\Sc|=a$ and $|\Sc\cap\Hc_{k}|=\rsf-1$  is $\binom{\rsf}{\rsf-1}\binom{\Hsf-\rsf}{a-\rsf+1}$. For each of thus sets $\Sc$, we want to choose one collection $\Qc$ 
where $|\Qc|=q$ and for each $\Yc\in \Qc$ we have $\Yc\subseteq \Sc$ and $\Yc\neq (\Sc\cap\Hc_{k})$.
It can be computed that the number of such collections for each set $\Sc$ is $\binom{\binom{a}{\rsf-1}-1}{q}$. Hence, the number of collections $\Qc$ in this group is $\Ysf_1$.

\item $|\Hc_k\cap \underset{\Yc\in \Qc}{\cup}\Yc|<\rsf-1$. The number of set $\Sc\subseteq [\Hsf]$ where $|\Sc|=a$ and $|\Sc\cap\Hc_{k}|<\rsf-1$  is $\binom{\Hsf}{a}-\binom{\Hsf-\rsf}{a-\rsf}-\binom{\rsf}{\rsf-1}\binom{\Hsf-\rsf}{a-\rsf+1}$. For each of thus sets $\Sc$, we want to choose one collection $\Qc$ 
where $|\Qc|=q$ and for each $\Yc\in \Qc$ we have $\Yc\subseteq \Sc$.
It can be computed that the number of such collections for each set $\Sc$ is $\binom{\binom{a}{\rsf-1}}{q}$. Hence, the number of collections $\Qc$ in this group is $\Zsf_1$.
\end{enumerate}
Hence, by  the inclusion-exclusion principle~\cite[Theorem~10.1]{combinatorics}, we prove Lemma~\ref{lem:needed memory size for improved scheme}.

\bibliographystyle{IEEEtran}
\bibliography{IEEEabrv,IEEEexample}
\end{document}